\documentclass[12pt,usenames,dvipsnames,a4paper]{article}
\pdfoutput=1

\usepackage{soul}

\usepackage[utf8]{inputenc}
\usepackage[T1]{fontenc}

\usepackage{changepage}
\usepackage{amsmath,amsfonts,amssymb}
\usepackage{epsf,amsmath,bbold,amsfonts,stmaryrd}
\usepackage{mathrsfs}
\usepackage{appendix}
\usepackage{caption}
\usepackage{cite}
\usepackage{color}
\usepackage{datetime}
\usepackage{float}
\usepackage{graphicx}
\usepackage[colorlinks]{hyperref}
\hypersetup{pageanchor=false,citecolor=red,urlcolor=red}
\usepackage{indentfirst}
\usepackage[numbers,square,comma,sort&compress,merge]{natbib} 
\usepackage{subfig}
\numberwithin{equation}{section}
\usepackage{mathtools}
\usepackage[colorinlistoftodos]{todonotes}
\usepackage{ytableau}
\usepackage{tabu}
\usepackage{esvect}

\usepackage{calc}

\allowdisplaybreaks

\def\f{\frac}

\def\l{\left}

\def\m{\mu}
\def\n{\nu}

\def\p{\partial}

\def\r{\right}

\def\x{\xi}

\def\be{\begin{equation}}
\def\ee{\end{equation}}

\def\bea{\begin{eqnarray}}
\def\eea{\end{eqnarray}}

\def\ba{\begin{array}}
\def\ea{\end{array}}

\def\bc{\begin{center}}
\def\ec{\end{center}}

\def\bl{\begin{flushleft}}
\def\el{\end{flushleft}}

\def\br{\begin{flushright}}
\def\er{\end{flushright}}

\def\bi{\begin{itemize}}
\def\ei{\end{itemize}}

\def\bt{\begin{tabular}}
\def\et{\end{tabular}}

\makeatletter

\newsavebox\myboxA
\newsavebox\myboxB
\newlength\mylenA

\newcommand*\xoverline[2][0.75]{%
    \sbox{\myboxA}{$\m@th#2$}%
    \setbox\myboxB\null
    \ht\myboxB=\ht\myboxA%
    \dp\myboxB=\dp\myboxA%
    \wd\myboxB=#1\wd\myboxA
    \sbox\myboxB{$\m@th\overline{\copy\myboxB}$}
    \setlength\mylenA{\the\wd\myboxA}
    \addtolength\mylenA{-\the\wd\myboxB}%
    \ifdim\wd\myboxB<\wd\myboxA%
       \rlap{\hskip 0.5\mylenA\usebox\myboxB}{\usebox\myboxA}%
    \else
        \hskip -0.5\mylenA\rlap{\usebox\myboxA}{\hskip 0.5\mylenA\usebox\myboxB}%
    \fi}
\makeatother

\newcommand{\ex}{\text{e}}

\def\be{\begin{equation}}
\def\ee{\end{equation}}

\def\bea{\begin{eqnarray}}
\def\eea{\end{eqnarray}}

\def\f{\frac}

\def\p{\partial}

\usepackage{xspace}
\newcommand*{\ie}{i.e., }
\newcommand*{\eg}{e.g., }

\newcommand*{\Eq}{Eq.\@\xspace}
\newcommand*{\Eqs}{Eqs.\@\xspace}

\usepackage{xifthen}
\newcommand*\diff{\mathrm{d}} 
\newcommand*\ldiff[2][]{ \ifthenelse{\isempty{#1}}{ \diff #2}{\diff^#1#2} \,} 
\let\limitint\int 
\renewcommand{\int}{\limitint \!} 

\begin{document}

\begin{titlepage}

\vspace*{-2.5cm}
\begin{adjustwidth}{}{-.45cm}
\br
{
\begin{tabular}{@{}l@{}}
\small LMU--ASC~12/22
 \end{tabular}
 }
\er
\end{adjustwidth}

\vspace*{1.5cm}
\begin{adjustwidth}{-1.3cm}{-.7cm}

\begin{center}
    \bf \Large{Field redefinitions, perturbative unitarity and Higgs inflation }
\end{center}
\end{adjustwidth}

\begin{center}
\textsc{Georgios K. Karananas,$^\star$~Mikhail Shaposhnikov,$^\dagger$~Sebastian Zell\,$^\dagger$}
\end{center}

\begin{center}
\it {$^\star$Arnold Sommerfeld Center\\
Ludwig-Maximilians-Universit\"at M\"unchen\\
Theresienstra{\ss}e 37, 80333 M\"unchen, Germany\\
\vspace{.4cm}
$^\dagger$Institute of Physics \\
Laboratory of Particle Physics and Cosmology\\ 
\'Ecole Polytechnique F\'ed\'erale de Lausanne (EPFL) \\ 
CH-1015 Lausanne, Switzerland
}
\end{center}

\begin{center}
\small
\texttt{\small georgios.karananas@physik.uni-muenchen.de}  \\
\texttt{\small mikhail.shaposhnikov@epfl.ch} 
\\
\texttt{\small sebastian.zell@epfl.ch} 
\end{center}

\begin{abstract}

For inflation driven by the Higgs field coupled non-minimally to gravity, we study the cutoff energy scale above which perturbation theory breaks down. Employing the metric formulation, we first give an overview of known results and then provide a new way to calculate a lower bound on the cutoff. Our approach neither relies on a gauge choice  nor does it require any calculation of amplitudes. Instead, it exploits the fact that the S-matrix is invariant under field redefinitions. In agreement with previous findings, we demonstrate that the cutoff is significantly higher during inflation than in vacuum, which
ensures the robustness of semi-classical predictions. Along the way, we generalize our findings to the Palatini formulation and comment on a useful parametrization of the Higgs doublet in both scenarios.

\end{abstract}

\end{titlepage}

\tableofcontents

\section{Introduction and motivation}

Given an effective field theory (EFT), it is important to determine in which interval of energies it is applicable. This energy range is bounded from above by the~\emph{cutoff scale $\Lambda$}, beyond which perturbation theory breaks down. In an S-matrix description, this means that perturbative unitarity is violated if the kinematical data of some of the quanta participating in scattering processes exceeds $\Lambda$.
Correspondingly, only phenomena with energies smaller than $\Lambda$ can be self-consistently described by the EFT. 

A straightforward way to gain information about the cutoff is  to perform  dimensional analysis.  Let $\tilde{\Lambda}$ be the smallest scale  by which higher-dimensional operators are suppressed in the Lagrangian of an EFT. Then a natural guess would be to identify $\Lambda$ with $\tilde{\Lambda}$. As a matter of fact, such a simple approach, which foregoes any calculation, is successful in many cases. However, there are exceptions to this rule. When determining the cutoff by an explicit S-matrix computation, it can happen that cancellations occur among different diagrams so that in the end amplitudes are independent of $\tilde{\Lambda}$. In such a case, the EFT remains valid for energies exceeding $\tilde{\Lambda}$. Therefore, $\tilde{\Lambda}$ in general only provides  an approximate lower bound on the cutoff scale, \ie $\tilde{\Lambda} \lesssim \Lambda$. 

Knowing $\Lambda$ is of particular significance for particle physics models aiming to capture the dynamics of the inflationary stage in the early Universe. Since gravity is involved, all these field theories are non-renormalizable EFTs for which $\Lambda \lesssim M_P$, where $M_P$ is the Planck mass. Nevertheless, they can be predictive and self-consistent, provided that all characteristic energies involved in physical processes are below the cutoff.

Among the plethora of inflationary models, the proposal~\cite{Bezrukov:2007ep} that the Higgs boson caused an early phase of exponential expansion stands out for two reasons. First, it does not require the introduction of new degrees of freedom beyond those already present in the Standard Model (SM) and gravity. This is very interesting since despite intensive searches no new particles have so far been detected. Secondly, the predictions of Higgs inflation (HI) are in excellent agreement with the latest observational data~\cite{Akrami:2018odb,BICEP:2021xfz}.

The phenomenological viability of HI requires the introduction of a large dimensionless coupling, $\xi \gg1$, between the Higgs doublet and the gravitational scalar curvature. This innocent-looking interaction of the Higgs field with gravity leads to a radical modification of the dynamics, especially at high energies. In particular, it lowers the cutoff scale of the system to a value significantly smaller than $M_P$. As discovered in~\cite{Burgess:2009ea,Barbon:2009ya}, the value of $\Lambda$  on top of the electroweak vacuum is 
\be
\label{eq:cutoffVacuum}
	\Lambda_{\rm EW} \sim \frac{M_P}{\xi} \ .
\ee
The Hubble parameter during HI is also of the order of $M_P/\xi$. Thus, if during inflation the cutoff scale were given by $\Lambda_{\rm EW}$, this would put into question the viability of the predictions derived from a semi-classical analysis.\footnote{See~\cite{Bezrukov:2007ep,Barvinsky:2008ia,Bezrukov:2008ej,DeSimone:2008ei,Burgess:2009ea,Barbon:2009ya,Bezrukov:2009db, Barvinsky:2009fy, Barvinsky:2009ii} for early investigations on the importance of quantum effects during HI.}

However, it was established~\cite{Bezrukov:2010jz} that the cutoff of the theory~\emph{depends on the background value of the Higgs} and becomes significantly larger than $\Lambda_{\rm EW}$ during inflation:
\be
\label{eq:cutoffInflation}
\Lambda_{\rm inf} \sim \frac{M_P}{\sqrt{\xi}} \ .
\ee
The fact that the above is higher than the characteristic energies guarantees the validity of the EFT and ensures the robustness of the  inflationary predictions.

The result~\eqref{eq:cutoffInflation} was derived in the unitary gauge, in which the Higgs doublet reduces to a real scalar field. With this choice of gauge, the dynamics leading to the cutoff of the theory reside in the massive gauge bosons of the SM. In the absence of gravity, the Higgs particle is the reason why the scattering of longitudinally polarized $W$ and $Z$ bosons do not violate perturbative unitarity---this is the very essence of the Higgs mechanism. Having modified the Higgs' dynamics by coupling it non-minimally to gravity, the aforementioned unitarization of amplitudes is no longer operative. As a result, the energy where HI cannot be trusted anymore is set by the effective inflationary masses of the gauge fields, which turn out to be proportional to~(\ref{eq:cutoffInflation}). 

Of course, any physical result cannot depend on the choice of the gauge. Thus, the estimate~(\ref{eq:cutoffInflation}) should also be valid if one did not insist on working in unitary gauge. In this case, the would-be Nambu-Goldstone (NG) bosons will be present in the action, making the scalar sector of HI nontrivial. Indeed, it was confirmed in~\cite{Ren:2014sya} that the interactions of the NG bosons also lead to the cutoff~(\ref{eq:cutoffInflation}). Since then, the validity of this result has been established using various methods~\cite{Fumagalli:2017cdo, Antoniadis:2021axu, Mikura:2021clt, Ito:2021ssc} (see also~\cite{Prokopec:2014iya, Escriva:2016cwl}).
 
The analysis in the presence of NG bosons leads to a puzzle. Namely, as was already noticed in~\cite{Lerner:2011it}, certain interactions between the Higgs and the NG fields appear suppressed by the low scale $M_P/\xi$. In the EFT language discussed above, this means that $\tilde{\Lambda} \sim  M_P/\xi$. On grounds of dimensional analysis, one might therefore be tempted to suspect that the value of the inflationary cutoff is lower than the one shown in~\Eq (\ref{eq:cutoffInflation}). Explicit studies of the relevant amplitudes show that this is not the case, see in particular~\cite{Ren:2014sya, Fumagalli:2017cdo, Antoniadis:2021axu, Ito:2021ssc}: The diagrams involving $M_P/\xi$ cancel, and the actual cutoff is identical to the result in unitary gauge.
 
The question arises if one can arrive at the conclusion that $\tilde{\Lambda} \sim M_P/\xi$ does not represent a cutoff for HI in a  simpler way than actually calculating amplitudes. As we shall show in the present paper, the answer is positive. Our method relies on the long-known fact that the S-matrix is invariant under certain classes of (local) field redefinitions~\cite{Chisholm:1961tha,Kamefuchi:1961sb}. This allows us to remove from the EFT all operators suppressed by the small scale $\tilde{\Lambda}$. Thus, one can view the seemingly problematic operators suppressed by it as an aftermath of not employing the most appropriate parametrization  for the degrees of freedom. In addition to making the analysis faster, the advantage of our approach is that it does not rely on nontrivial cancellations between  amplitudes, which makes it more robust against computational mistakes.\footnote{In fact, part of the motivation for our present study comes from a---meanwhile rectified---issue in the calculation of amplitudes in an earlier version of~\cite{Ito:2021ssc}.}

 A comment is in order about what happens above the cutoff scale, \ie what are the possible  UV-completions of HI. Certainly a possibility is to integrate in new heavy degrees of freedom that are assumed to live in the proximity of the cutoff scale and make the theory weakly coupled. Corresponding models for HI have been constructed \eg in~\cite{Giudice:2010ka, Gorbunov:2013dqa, Barbon:2015fla, Ema:2017rqn}. In all these theories, however, only the low vacuum cutoff~\eqref{eq:cutoffVacuum} is relevant, \ie the new heavy particles already enter at the scale $M_P/\xi$, independently of the background value of the Higgs field. Therefore, the additional degrees of freedom can leave their imprints on inflationary dynamics and  influence predictions. It is conceivable that all such weakly coupled UV-completions exhibit this property,\footnote{We thank José Espinosa for a comment about this point.} although this remains to be proved.

The above is not the only option though. It may well happen that HI UV-completes in a different manner, for instance by ``self-healing''~\cite{Bezrukov:2010jz, Aydemir:2012nz}, or by ``classicalizing''~\cite{Dvali:2010bf, Dvali:2010jz, Dvali:2011th}. If this happens, the particles that already exist in the infrared domain enter into a regime of strong coupling around the cutoff scale. This option makes it possible that during HI the effects of new physics only become relevant above the high energy $M_P/\sqrt{\xi}$ shown in~\Eq~\eqref{eq:cutoffInflation}~\cite{Bezrukov:2010jz, Bezrukov:2014ipa}. Consequently, the inflationary predictions as derived in~\cite{Bezrukov:2007ep} remain robust.

Finally, it is important to remark that all the previous discussion referred to HI in the metric formulation of gravity, as it was originally proposed. However, gravity exists in several other incarnations. They are all indistinguishable as far as purely gravitational interactions are concerned, but can lead to distinct observable predictions once matter is included. For example, including a non-minimal coupling to the scalar curvature breaks the equivalence between the various versions of General Relativity. Therefore, HI is sensitive to the choice of gravitational formulation---several options have been explored so far~\cite{Bauer:2008zj,Rasanen:2018ihz, Raatikainen:2019qey, Langvik:2020nrs, Shaposhnikov:2020gts}.  Which version of gravity is used also finds its way into the cutoff. In particular, the Palatini scenario of HI leads to a higher cutoff scale on top of the electroweak vacuum~\cite{Bauer:2010jg}, which does not increase significantly during inflation~\cite{Shaposhnikov:2020geh}. Therefore, there is no question about a potentially problematic low value of the perturbative cutoff in Palatini Higgs inflation. 

This article is organized as follows. In Sec.~\ref{sec:toyModel}, we construct a simple toy model that captures the essence of what happens in Higgs inflation. We choose an extreme case, in which a seemingly non-renormalizable theory is in fact a free theory in disguise. Subsequently, we turn to the actual Higgs inflation in Sec.~\ref{sec:Higgs_Inflation}. First, we briefly introduce the model and discuss the situation in the unitary gauge. Then, we use a Cartesian parametrization for the would-be NG modes in the Higgs doublet. We find a simple change of variables that eliminates the scale $M_P/\xi$ from the action, in complete analogy with the toy model. Subsequently, we show that by using a different, exponential, parametrization for the NG modes the resulting action is liberated from artifacts of a poor choice of variables. Thus,  reading the cutoff scale during inflation becomes straightforward. In Sec.~\ref{sec:PalatiniHiggs}, we briefly discuss  the different parametrizations of the Higgs doublet for the Palatini version of Higgs inflation. We conclude in Sec.~\ref{sec:conclusion}.

\section{A toy model}
\label{sec:toyModel}

 Before turning to HI, we think that it is beneficial for the reader to make our point in a simple toy model. We take two real scalar fields $\varphi_1$ and $\varphi_2$ with the following action in a four-dimensional Minkowski spacetime\,\footnote{In our conventions $\eta_{\mu\nu}={\rm diag}(-1,+1,+1,+1)$.} 
\be 
\label{toyModel}
S = - \frac{1}{2}\int\diff^4x \left[(\partial_\mu \varphi_1)^2   +(\partial_\mu \varphi_2)^2  + \frac{2\varphi_2}{\tilde{\Lambda}} \partial_\mu  \varphi_1 \partial^\mu  \varphi_2 +\frac{c}{ \tilde{\Lambda}^2} \varphi_2^2 \partial_\mu \varphi_2 \partial^\mu \varphi_2 \right] \ ,
\ee
where $\tilde{\Lambda}$ carries dimension of mass and $c$ is a real parameter. At first sight, one could think that the above describes two fields interacting nontrivially via derivative mixings. Furthermore, one might suspect that it is only valid up to energies of the order of $\tilde \Lambda$, \ie that this scale represents the cutoff of the theory.

As a first step toward determining  the true nature of the energy scale $\tilde \Lambda$, let us calculate an explicit amplitude. We choose the process $2 \varphi_1 \rightarrow 2 \varphi_2$. To lowest order in  $1/\tilde{\Lambda}^2$, only two diagrams contribute (one internal $\varphi_2$-line, in $t$- and $u$-channels). The different contributions cancel precisely and the resulting amplitude vanishes\,\footnote{In the computation, this finding is due to the fact that $s+t+u=0$ in a massless theory, where $s$, $t$ and $u$ are the Mandelstam variables.}
\begin{equation}
	\mathcal{A}_{2 \varphi_1 \rightarrow 2 \varphi_2} = 0 \;.
\end{equation}
 Thus, $\varphi_1$ and $\varphi_2$ do not interact at the lowest order in $\tilde{\Lambda}$ and this process is independent of $\tilde \Lambda$. 
Nevertheless, to make a definite statement about the dynamics of the toy-model, one would need to compute all possible amplitudes to all orders, a rather tedious study.

There is an easier way to determine whether there is interaction in the theory~\eqref{toyModel}. It does not require any calculation of an amplitude. First we note that we can rearrange the terms to rewrite the action as
\be
S= -\f 1 2\int\diff^4x\left[\left(\p_\m \varphi_1 +\f{\varphi_2}{\tilde \Lambda}\p_\m \varphi_2 \right)^2 +\left(1+(c-1)\f{\varphi_2^2}{\tilde \Lambda^2}\right)(\p_\m \varphi_2)^2 \right] \ . 
\ee 
Then we introduce two new fields $\chi_1$ and $\chi_2$, given by
\be
\label{fieldRedefinition1}
	\chi_1=  \varphi_1 + \frac{\varphi_2^2}{2 \tilde{\Lambda}} \ ,
\end{equation}
and 
\bea \label{fieldRedefinition2}
&\chi_2 ={\displaystyle \limitint^{\varphi_2}}\diff\varphi\sqrt{1 + (c-1)\frac{\varphi^2}{\tilde{\Lambda^2}}} =\f{\varphi_2}{2}\sqrt{1 + (c-1)\frac{\varphi_2^2}{\tilde{\Lambda^2}}} + \frac{\tilde{\Lambda}}{2\sqrt{(c-1)}}\ {\sinh}^{-1}\left( \frac{\sqrt{(c-1)}\varphi_2}{\tilde{\Lambda}}\right)  \ ,&
\eea
where we assumed $c>1$. The field variables $\chi_1$ and $\chi_2$ make apparent that the toy model~\eqref{toyModel} actually describes two completely decoupled free massless scalar fields,
\begin{equation} 
S = -\frac{1}{2}\int \diff^4x\left[ (\partial_\mu \chi_1)^2 + (\partial_\mu \chi_2)^2 \right] \ ,
\end{equation}
albeit initially written in a peculiar manner. Here we exploit the fact that the S-matrix is invariant under local field redefinitions that leave the quadratic Lagrangian invariant \cite{Chisholm:1961tha, Kamefuchi:1961sb}. Since the transformations \eqref{fieldRedefinition1} and \eqref{fieldRedefinition2} fulfill these properties, it follows that all S-matrix elements of the initial theory \eqref{toyModel} vanish, \ie also $\varphi_1$ and $\varphi_2$ are free fields. We thus conclude that $\tilde \Lambda$ does not have any physical meaning but rather is an artifact of a poor choice of variables. 

We shall describe yet another way to derive this result. It relies on studying the field-derivative manifold. If it is flat, this guarantees the existence of appropriate field redefinitions such that the kinetic terms of both fields become canonical  \cite{Cornwall:1974km} (see also e.g.~\cite{Garcia-Bellido:2011kqb,Karananas:2016kyt}).
 To explain what we mean in more details, consider the following more general theory, again in 4 spacetime dimensions  
\be
\label{eq:sigma_model_form}
S = -\f 1 2 \int \diff^4x \, G_{IJ}\p_\mu\Phi_I\p^\mu \Phi_J \ ,
\ee
where $\Phi_I=(\varphi_1,\varphi_2)$, $I,J=1,2$ and $G_{IJ}=G_{IJ}(\varphi_2)$ is the metric of the target space that in our considerations depends on $\varphi_2$ only; summation over all repeated indexes is understood.
Being two-dimensional in the internal space, all features of the field-derivative manifold are captured by the scalar curvature $\kappa$.\footnote{Remember that in two dimensions the full curvature tensor reads
\be
\label{eq:curv_tens}
\kappa_{IJKL}= \f{\kappa}{2}  (G_{IK}G_{JL}-G_{JK}G_{IL}) \ .
\ee} It is given by
\be
\label{eq:curv_2d}
\kappa = \f{G_{11}'F'-2FG_{11}''}{2F^2} \ ,~~~F = G_{11}G_{22}-G_{12}^2\ ,
\ee
where prime denotes derivative with respect to $\varphi_2$. 
Provided that the components of the internal metric are such that $\kappa=0$ in~\eqref{eq:curv_2d}, the theory under consideration represents two free scalar fields in disguise. Viewed from this perspective, it is  obvious that because $G_{11}$ is constant, our toy model~(\ref{toyModel}) is a free field theory. 

Finally, we remark that the situation for internal spaces with dimension $n>2$ is similar, nonetheless slightly more involved. There, the relation~(\ref{eq:curv_tens}) does not hold anymore, meaning that the flatness of the target manifold requires that either the Ricci tensor is zero when $n=3$, or that all the $n^2(n^2-1)/12$ components of the curvature tensor vanish for $n\ge 4$. We will discuss such a situation later in Sec.~\ref{ssec:CartesianAnalysis}, where we study Higgs inflation.

\section{Metric Higgs inflation}
\label{sec:Higgs_Inflation}

\subsection{Generalities}
\label{ssec:Generalities}

We consider the relevant parts of the SM Higgs sector coupled non-minimally to gravity in the metric formulation. Our starting point is the action 
\begin{equation} 
\label{actionHiggsJordan}
	 S = \int \diff^4 x \sqrt{g} \Bigg[ \left( \frac{M_P^2}{2} + \xi H^\dagger H \right) R - g^{\m\n} \left( \mathcal D_\m H \right)^{\dagger}  \mathcal D_\n H - V(H) \Bigg] + S_{\rm gauge}\ ,
\end{equation}
where $g={\rm det}(-g_{\m\n})$, $R$ is the scalar curvature, $H$ the  Higgs doublet, and $\mathcal D_\mu$ corresponds to the gauge covariant derivative 
\be
\mathcal D_\m H = \p_\m H - i \f {g_2}{2} A_\m H - i \f {g_1}{2} B_\m H \ .
\ee
Here 
 \be
A_\m =   \begin{pmatrix}
	A_\m ^3  & A_\m^1-i A_\m^2 \\
	A_\m^1+i A_\m^2 & -A_\m ^3  
	\end{pmatrix} \ ,
\ee
and $B_\m$ represent the gauge fields of the $SU(2)_L$ and $U(1)_Y$ groups, and $g_2$ and $g_1$ denote the respective couplings.  Moreover, $V(H)$ is the usual Higgs potential, the form of which at high energies is
\begin{equation}
	V(H) \approx \lambda (H^\dagger H)^2 \ ,
\end{equation} 
with $\lambda$ setting the strength of the field's self-interaction. Finally, 
\be
\label{eq:gauge_act_kin}
S_{\rm gauge} = \int \diff^4 x \sqrt{g}\left( -\f{1}{8} {\rm tr}(F_{\m\n}^2) -\f{1}{4} B_{\m\n}^2 \right) \ ,
\ee
where the field strengths read as usual
\bea
F_{\m\n} = \p_\m A_\n -\p_\n A_\m -i \f{g_2}{2} [A_\m,A_\nu] \ ,~~~B_{\m\n} = \p_\m B_\n -\p_\n B_\m \ , 
\eea
with the square brackets denoting the commutator.  Although there is no difficulty in working in the Jordan frame, we will instead move to the Einstein frame, where gravity is canonical. To do that we Weyl-rescale the metric as follows
\begin{equation} 
\label{eq:WeylTrafo}
	g_{\mu\nu} \rightarrow \Omega^{-2} 	g_{\mu\nu} \ ,
\end{equation}
where 
\begin{equation} \label{omega}
	\Omega^2 =1+ 	\frac{2\xi H^\dagger H}{M_P^2} \ ,
\end{equation}
is the conformal factor. Standard manipulations lead to the Einstein frame action
\bea
\label{eq:actionHiggsEinstein}
&&S = \int \diff^4 x \sqrt{g} \Bigg[ \frac{M_P^2}{2} R - \frac{1}{\Omega^2} g^{\m\n} \left( D_\m H \right)^{\dagger}  D_\n H \nonumber \\
&&\qquad\qquad\qquad\quad- \frac{3 \xi^{2}}{M_P^2\Omega^4} g^{\m\n} \partial_\m (H^{\dagger}H) \partial_\n (H^{\dagger}H) - U(H)\Bigg]+S_{\rm gauge} \ ,
\eea
with
\be
\label{eq:eins_pot}
U(H)=\frac{\lambda (H^\dagger H)^2}{\Omega^4} \ .
\ee
We note that the field strengths of the gauge fields are not affected by the Weyl transformation, \ie $S_{\rm gauge}$ is still given by \Eq \eqref{eq:gauge_act_kin}. In contrast, the kinetic term for $H$ and its interactions with the gauge fields---both contained in the covariant derivative---become noncanonical, there is a dimension-six operator because of the inhomogeneous transformation of the scalar curvature under~(\ref{eq:WeylTrafo}), and also the potential gets rescaled.

Going to the Einstein frame simplifies a bit the considerations, since the interactions involving gravitons are Planck suppressed and thus can be safely neglected.\footnote{Had we been working in the Jordan frame instead, we would have to account for the explicit kinetic mixing between the gravitons and Higgs, see \eg\cite{Bezrukov:2010jz}, as well as~\cite{Bezrukov:2012hx} where the unitarity of the closely related Higgs-dilaton inflation~\cite{Shaposhnikov:2008xb,Garcia-Bellido:2011kqb} was analyzed.} All the following considerations take place in flat Minkowski spacetime, a good approximation for the energies we are interested in. In addition, we assume that the background value $\bar h$ of the Higgs field is constant. 

\subsection{Analysis in unitary gauge}
\label{ssec:Unitary_Analysis}

Typically HI is studied in unitary gauge since this simplifies the theory considerably. In the scalar sector only the physical excitation of the Higgs field is present, so in what follows we take $H=\l(0,\mathcal H/\sqrt{2}\r)^T$, where $\mathcal H$ is a real scalar field. The fact that the dynamics eventually boil down to the one of a single field may give the (inaccurate) impression that important information is completely erased, especially when it comes to determining the cutoff. This is not correct: One needs to look elsewhere since the problem propagates  all the way to the gauge sector of the theory. The reason is simple and is actually the whole essence of the Higgs mechanism.

In unitary gauge, each of the massive gauge bosons of the SM propagates three degrees of freedom, two transverse and one longitudinal. The latter are actually the would-be NG bosons contained in the Higgs doublet. At high energies, 2$\,\to\,$2 scattering processes with these components as external states dominate and are responsible for the divergent behavior of the corresponding amplitudes. They are proportional to $(E/m_V)^2$, where $E$ represents the typical energy of the process and $m_V$ is the mass of the corresponding vector.
In the SM, this issue is resolved after taking into account processes in which the Higgs excitation is exchanged. Of course, this is not an accident since the interactions of the gauge bosons among themselves as well as with $\mathcal H$ are nontrivially related to each other.  More explicitly, the coupling of the Higgs to the vector bosons is proportional (up to the gauge coupling) to $m_V$. Only then can the amplitudes associated with such processes interfere destructively, their growth with energy  be cancelled out and the result become finite. 

Let us now discuss how this picture changes, dramatically, in HI. Because of the strong non-minimal interaction with gravity, the cancellations between dangerous diagrams no longer take place and amplitudes for scattering of massive gauge bosons grow unboundedly with energy. In order to demonstrate how this comes about, we drop the potential $U$, the contribution of which is subdominant as we show at the end of this section,  and concentrate on the kinetic term of the Higgs and the gauge sector. The  action~(\ref{eq:actionHiggsEinstein}) in the unitary gauge becomes 
\be
\label{eq:action_unitary}
S=-\f 1 2\int\diff^4x \left[ \f{1}{\Omega^2}\left(1 + \f{6\x^2}{M_P^2}\f{\mathcal H^2}{\Omega^2}\right)(\p_\m \mathcal H)^2 + \f{1}{4}{\rm tr}(F_{\m\n}^2) +\f{1}{2} B_{\m\n}^2+ \f{\mathcal H^2}{8\Omega^2}{\rm tr}(V_\m^2) \right] \ ,
\ee
where \Eq \eqref{omega} implies that
\begin{equation} \label{omegaUnitary}
	\Omega^2 =1+ 	\frac{\xi\mathcal H^2}{M_P^2} \ ,
\end{equation}
and we introduced~\cite{Bezrukov:2009db}
\be
V_\m =   \begin{pmatrix}
	g_2 A_\m ^3 -g_1 B_\m & g_2(A_\m^1-i A_\m^2) \\
	g_2(A_\m^1+i A_\m^2) & -(g_2 A_\m ^3 - g_1 B_\m ) 
	\end{pmatrix} \ .
\ee

Now we consider excitations $h$ on top of the  background value $\bar h$, \ie we split $\mathcal H=\bar h+h$ and expand in powers of $h$. Plugging this into the action \eqref{eq:action_unitary} and keeping the leading terms which are at most quadratic  in derivatives and linear in the Higgs excitation, we find
\begin{equation}
    S \simeq -\f 1 2\int\diff^4x\left[ \f{6 M_P^2}{\bar h^2}(\p_\m h)^2 +\f{1}{4}{\rm tr}(F_{\m\n}^2)+\f{1}{2} B_{\m\n}^2 + \f{M_P^2}{8\xi}{\rm tr}(V_\m^2) +\f{M_P^4}{4\xi^2\bar h^3}h\,{\rm tr}(V_\m^2)\right] \ ,
\end{equation} 
where we used that the background field during inflation satisfies $\bar h\gg M_P/\sqrt{\xi}$.
Let us now normalize canonically the kinetic term by introducing
\be
\chi = \f{\sqrt 6 M_P}{\bar h}h  \ ,
\ee
in terms of which the above becomes
\be
\label{eq:action_unitary_canonical}
S \simeq -\f 1 2 \int\diff^4x\left[ (\p_\m \chi)^2 + \f{1}{4}{\rm tr}(F_{\m\n}^2) +\f{1}{2} B_{\m\n}^2+ \f{M_P^2}{8\xi}{\rm tr}(V_\m^2) + \f{M_P^3}{4\sqrt{6}\xi^2\bar h^2}\chi\,{\rm tr}(V_\m^2)\right] \ .
\ee
From this expression we can easily understand the essence of the situation. First of all, we notice that in the inflationary background $\bar h\gg M_P/\sqrt{\xi}$, the vector fields acquire a large effective mass $m_V \propto M_P/\sqrt{\xi}$. At the same time, the relevant interaction between the scalar excitations and the gauge bosons (last term in \Eq~\eqref{eq:action_unitary_canonical}) is suppressed due to inverse powers of $\bar h$ and vanishes in the limit $\bar h \rightarrow \infty$.     The fact that it is no longer simply proportional to the masses of the gauge bosons is the reason why the delicate cancellation that tames the divergent amplitudes associated with the longitudinal gauge bosons cannot take place in HI. As we already mentioned, the main point is that the kinetic and self-interaction terms of the gauge fields contained in ${\rm tr}(F_{\m\n}^2)$ and $B_{\m\n}^2$ are not altered when moving to the Einstein frame. In contrast, both their masses and interactions with the excitations of the Higgs change after Weyl-rescaling the metric.

It is straightforward to infer the inflationary cutoff scale from the above considerations. Since the scattering amplitudes  among the massive gauge bosons are inversely proportional to the effective mass $m_V \propto M_P/\sqrt{\xi}$ and can no longer be cancelled by interactions with the Higgs field, perturbation theory breaks down for energies exceeding $m_V$. This leads to the cutoff scale~\cite{Bezrukov:2010jz}
\be 
\label{eq:cutoffInflationUnitary}
\Lambda_{\rm inf} \sim \frac{M_P}{\sqrt{\xi}} \ ,
\ee
in accordance with \Eq~\eqref{eq:cutoffInflation}. We remark that it is easy to generalize this analysis in unitary gauge to different formulations of GR, such as for instance Palatini gravity. In this case, the masses of the gauge bosons are also proportional to $M_P/\sqrt{\xi}$, but what changes is the interaction of the Higgs excitation  with the latter. Although the suppression is weaker than in the metric case, it is still strong enough to nullify the mechanism responsible for the cancellations of dangerous amplitudes. Thus, the scale of unitarity violation during inflation is also given by the effective mass of the gauge bosons, and the inflationary cutoff scale shown in \Eq~\eqref{eq:cutoffInflationUnitary} equally applies to Palatini Higgs inflation~\cite{Shaposhnikov:2020geh}.

Finally, let us also briefly explain what happens with the contributions coming from the potential. A straightforward computation following \cite{Bezrukov:2010jz} reveals that once~(\ref{eq:eins_pot}) is expanded around a fixed background, we find 
\be
U(H)\simeq \f{\lambda M_P^4}{4\xi^2}\left[1+ \f{M_P^2}{\xi \bar h^2}\sum_{n=1}c_n \left(\f{h}{\bar h}\right)^n\right]\simeq \f{\lambda M_P^4}{4\xi^2}\left[1+ \f{M_P^2}{\xi \bar h^2}\sum_{n=1}\tilde{c}_n \left(\f{\chi}{M_P}\right)^n\right] \ ,
\ee
with $c_n$ , $\tilde{c}_n$  numerical factors $\mathcal O(1)$. We thus see that the terms coming from the potential are heavily suppressed. Actually, the suppression is exponential as a result of the exponential map between $\bar h$ and the canonically normalized background field.

Before moving on, it is important to reiterate the main point of the above analysis. Choosing to work in the unitary gauge is a matter of convenience and cannot change the physics. Any self-consistency issue cannot disappear in another gauge---it ought reemerge, albeit potentially in a different sector of the theory. 

\subsection{Analysis in Cartesian coordinates}
\label{ssec:CartesianAnalysis}

Next, we shall derive the inflationary cutoff scale without working in the unitary gauge. Hence, we need only focus on the scalar sector of the theory and  can drop the gauge bosons from the covariant derivative. For the reasons explained above, we also do not take into account the potential. Our starting point is thus
\be
\label{eq:action_starting}
S= -\int \diff^4x \f{1}{\Omega^2}\left[\p_\mu H^\dagger\p^\mu H +\frac{3 \xi^{2}}{M_P^2\Omega^2} \partial_\mu (H^{\dagger}H) \partial^\mu (H^{\dagger}H)\right] \ . 
\ee
We parametrize the Higgs doublet as 
\begin{equation} 
\label{doubletCartesian}
	H =
	\frac{1}{\sqrt 2}\begin{pmatrix}
	    \pi_1+i\pi_2\\
		\bar{h} + h + i \pi_3 
	\end{pmatrix} \ ,
\end{equation}
where as before $h$ is the physical Higgs field and the $\pi_a$'s, $a=1,2,3$, are real scalars---the would-be NG modes. This parametrization coincides with the one used in \cite{Ito:2021ssc}, once $\pi^1$ and $\pi^2$ are combined into the complex field $\pi^+ = \f{1}{\sqrt 2}(\pi^1+i\pi^2)$.

Plugging the split \eqref{doubletCartesian} into the action~\eqref{eq:action_starting}, we obtain to fourth order in perturbations
\begin{equation}
\label{eq:action_perts_cart}
    S= \int \diff^4x \Big(\mathcal L_2+\mathcal L_3+\mathcal L_4 \Big) \ ,
\end{equation}
where
\bea
\label{eq:L2}
&\mathcal L_2=-\f{1}{2\bar \Omega^2} \left[\left(1+\f{6\xi^2\bar h^2}{M_P^2\bar\Omega^2}\right) (\p_\mu h)^2+(\p_\mu\pi_a)^2\right] \ ,\qquad\qquad\qquad\qquad\qquad~&\\
\label{eq:L3}
&\mathcal L_3= \f{\xi\bar h}{M_P^2\bar \Omega^4}\left[\left( 1-\f{6\xi}{\bar\Omega^2}\left(1-\f{\xi\bar h^2}{M_P^2}\right) \right)h(\p_\mu h)^2-3\xi \p_\mu h\p^\mu\pi_a^2 +h (\p_\mu \pi_a)^2 \right] \ ,&\\
&\mathcal L_4=-\f{\xi}{2M_P^2\bar \Omega^6}\Bigg[  \left(3(1+6\xi)\bar \Omega^2-4(1+21\xi)+\f{72\xi}{\bar \Omega^2}\right)h^2(\p_\mu h)^2 \nonumber\qquad\qquad~&\\
\label{eq:L4}
&~~~~+\left(1-\f{3\xi\bar h^2}{M_P^2} \right)\left(3\xi\p_\mu h^2\p^\mu \pi_a^2-h^2 (\p_\mu \pi_a)^2\right)-\left(\bar\Omega^2+\f{12\xi^2\bar h^2}{M_P^2}\right)(\p_\mu h)^2\pi_a^2 \nonumber& \\
&\qquad\qquad\qquad\qquad\qquad\qquad\qquad\qquad\qquad\quad~-\bar\Omega^2\left(\pi_a^2 (\p_\mu\pi_b)^2-\f{3\xi}{2}(\p_\mu\pi_a^2)^2\right)\Bigg] \ ,&
\eea
 are the quadratic, cubic and quartic contributions, respectively. Summation over repeated indices is understood and we defined (in analogy to \Eq \eqref{omegaUnitary})
\be
\label{eq:barOmega}
\bar \Omega = \sqrt{1+\f{\xi\bar h^2}{M_P^2}} \ .
\ee
We now make the kinetic terms in $\mathcal L_2$ canonical. This is achieved by introducing the fields $\chi$ and $\sigma_a$, related to $h$ and $\pi_a$ via the rescaling 
\be
\label{eq:field_resc}
h=\f{\bar\Omega}{\sqrt{1+\f{6\xi^2\bar h^2}{M_P^2\bar\Omega^2}}}\chi \ ,~~~\pi_a= \bar \Omega \sigma_a \ ,
\ee
respectively. It is easy to see that in terms of the canonically normalized fields, the expressions~(\ref{eq:L2})-(\ref{eq:L4}) become
\bea
\label{eq:can_kin_L2}
&\mathcal L_2=-\f{1}{2}\left[(\p_\mu \chi)^2+(\p_\mu\sigma_a)^2\right] \ ,\qquad\qquad\qquad\qquad\qquad\qquad\qquad\qquad\qquad\qquad& \\
&\mathcal L_3=\f{\xi\bar h}{M_P^2\sqrt{\bar\Omega^2+\f{6\xi^2\bar h^2}{M_P^2}}}\left[\left(1-\f{6\xi}{\bar\Omega^2+\f{6\xi^2\bar h^2}{M_P^2}}\right)\chi(\p_\mu \chi)^2-3\xi\p_\mu \chi \p^\mu \sigma_a^2 +\chi(\p_\mu\sigma_a)^2\right] \ ,\quad& \\
&\mathcal L_4=-\f{\xi}{2M_P^2} \Bigg[
\f{3(1+6\xi)\bar \Omega^4-4(1+21\xi)\bar \Omega^2+72\xi}{\left(\bar \Omega^2+\f{6\xi^2\bar h^2}{M_P^2}\right)^2}\chi^2(\p_\mu \chi)^2+\f{1-\f{3\xi\bar h^2}{M_P^2}}{\bar \Omega^2+\f{6\xi^2\bar h^2}{M_P^2}}\Big(3\xi\p_\mu \chi^2\p^\mu \sigma_a^2\nonumber\qquad\quad&\\
&\quad\quad\quad~-\chi^2 (\p_\mu \sigma_a)^2\Big)-\f{\bar\Omega^2+\f{12\xi^2\bar h^2}{M_P^2}}{\bar \Omega^2+\f{6\xi^2\bar h^2}{M_P^2}}(\p_\mu\chi)^2\sigma_a^2-\sigma_a^2 (\p_\mu\sigma_b)^2+\f{3\xi}{2}(\p_\mu\sigma_a^2)^2\Bigg] \ .&
\eea
This action is equivalent to the one derived in \cite{Ito:2021ssc}.

Let us now take $\bar h\gg M_P/\sqrt{\xi}$, corresponding to the inflationary energy domain. In this limit, the kinetic terms remain unchanged, see \Eq~(\ref{eq:can_kin_L2}), while the interaction terms simplify considerably. The action becomes 
\bea 
\label{actionHiggsLeading}
&S\simeq  -\f 1 2\int\diff^4x \Bigg[(\p_\mu \chi)^2+(\p_\mu\sigma_a)^2+ \f{\xi}{M_P}\sqrt{6}\p_\mu\chi\p^\mu\sigma_a^2+\f{\xi^2}{M_P^2}\f 3 2  (\p_\mu\sigma_a^2)^2\nonumber&\\
&\qquad-\f{\xi}{M_P^2}\left(2 (\p_\mu\chi)^2\sigma_a^2 + \f 3 2 \p_\mu\chi^2\p^\mu\sigma_a^2+\sigma_a^2 (\p_\mu\sigma_b)^2\right)\nonumber&\\
&\qquad-\f{1}{M_P}\left(\f{1}{2 \sqrt{6}}\p_\mu \chi  \p^\mu\sigma_a^2 +\sqrt{\f{2}{3}}\chi\left( (\p_\mu\chi)^2 + (\p_\mu\sigma_a)^2\right)\right)\nonumber&\\
&\qquad\qquad+\f{1}{M_P^2}\left(\f{1}{4} \p_\mu\chi^2\p^\mu\sigma_a^2 +\f{1}{6} (\p_\mu\chi)^2\sigma_a^2+ \f{1}{2}\chi^2\left( (\p_\mu\chi)^2 + (\p_\mu\sigma_a)^2\right)\right) +\mathcal O(\xi^{-1})\Bigg] \ ,&
\eea
where we also used that $\xi \gg 1$. We immediately notice that there are three different energy scales:  $M_P/\xi\ll M_P/\sqrt{\xi}\ll M_P$. Therefore, the only statement about the cutoff $\Lambda_{\rm inf}$ that we can make at this point without calculating amplitudes is
\begin{equation} 
\label{cutoffBoundCartesian}
	\Lambda_{\rm inf} \gtrsim \frac{M_P}{\xi} \ .
\end{equation}
We are interested in understanding whether $M_P/\xi$ actually represents a cutoff, or if it is possible to derive a stronger bound than the above. To this end, we can proceed in analogy with the two-field toy model that we presented in the previous section. Namely, we notice that the kinetic term for the physical Higgs together with the seemingly problematic operators suppressed by the low scale $M_P/\xi$ combine neatly in a perfect square:
\bea 
\label{actionHiggsLeading2}
&S \simeq-\f 1 2 {\displaystyle\int} \diff^4x \Bigg[\left(\p_\mu\chi+\sqrt{\f{3}{2}}\frac{\xi}{M_P}\p_\mu\sigma_a^2\right)^2 + (\p_\mu\sigma_a)^2&\nonumber\\
&-\f{\xi}{M_P^2}\bigg(2 (\p_\mu\chi)^2\sigma_a^2 + \f 3 2 \p_\mu\chi^2\p^\mu\sigma_a^2+\sigma_a^2 (\p_\mu\sigma_b)^2\bigg) &\nonumber\\
&-\f{1}{M_P}\left(\f{1}{2 \sqrt{6}}\p_\mu \chi  \p^\mu\sigma_a^2 +\sqrt{\f{2}{3}}\chi\left( (\p_\mu\chi)^2 + (\p_\mu\sigma_a)^2\right)\right)\nonumber\\
&+\f{1}{M_P^2}\left(\f{1}{4} \p_\mu\chi^2\p^\mu\sigma_a^2 +\f{1}{6} (\p_\mu\chi)^2\sigma_a^2+ \f{1}{2}\chi^2\left( (\p_\mu\chi)^2 + (\p_\mu\sigma_a)^2\right)\right) +\mathcal O(\xi^{-1})\Bigg]\ .& 
\eea
 The form of the above written this way is highly suggestive---it dictates that we introduce
\be \label{fieldRedefinition}
\chi= \rho-\sqrt{\f{3}{2}}\frac{\xi}{M_P}\sigma_a^2 \ .
\ee
This field transformation is fully analogous to the study of the toy model (see \Eq~\eqref{fieldRedefinition1}), after the identifications $\chi \leftrightarrow \varphi_1$, $\rho \leftrightarrow \chi_1$, $\sigma_a \leftrightarrow \varphi_2$ and $M_P/(\sqrt{6} \xi) \leftrightarrow \tilde{\Lambda}$.
Plugging \Eq \eqref{fieldRedefinition} into the action~\eqref{actionHiggsLeading2} and keeping again terms which are at most quartic in the fields, we obtain (after an integration by parts) 
\bea
\label{eq:compare_appendix}
&&S \simeq-\f 1 2 \int \diff^4x \Bigg[\left(\p_\mu\rho\right)^2 + (\p_\mu\sigma_a)^2+\f{\xi}{M_P^2}\left(\sigma_a^2\rho\square \rho+\f 1 4 (\p_\m\sigma_a^2)^2\right) +\xi\,\mathcal O(\xi^{-1})\Bigg] \ ,
\eea
where we only kept the leading terms.
This shows that the scale $M_P/\xi$ is spurious, and the inflationary cutoff of the theory obeys
\be
\label{eq:true_cutoff}
\Lambda_{\rm inf} \gtrsim \f{M_P}{\sqrt{\xi}} \ . 
\ee
Without any calculation of an amplitude, we have demonstrated that the lowest scale that can appear in interactions involving the Higgs excitations and the would-be NG bosons is $M_P/\sqrt{\xi}$. This conclusion agrees with explicit studies of such scattering processes \cite{Ren:2014sya, Fumagalli:2017cdo, Antoniadis:2021axu, Ito:2021ssc}, which moreover show that the bound~\Eq~\eqref{eq:true_cutoff} is sharp, \ie the inflationary cutoff actually is $\Lambda_{\rm inf} \sim M_P/\sqrt{\xi}$ (see \Eq~\eqref{eq:cutoffInflation}).

The reader is welcome to check that the above analysis can be extended to higher orders in perturbations. We note, though, that in this case operators suppressed by scales involving other fractional powers of $\xi$ may appear; again, these are spurious and only an artifact of making a field redefinition which is quadratic in the fields. To accurately capture higher orders in perturbations effects, we need to go to higher in perturbations in the field redefinition. This is certainly doable, although tedious. Instead of that, we will subsequently demonstrate a cleaner way to proceed, which employs an exponential representation for the Higgs doublet. 

\subsection{Field space curvature}
Before that, we shall make a short side remark about a complementary way to analyze the action \eqref{actionHiggsLeading}. It relies on studying its field-space curvature, which we discussed before using the toy model of Sec.~\ref{sec:toyModel}. This approach was already applied to Higgs inflation in \cite{Mikura:2021clt} (see also \cite{Alonso:2015fsp, Nagai:2019tgi}).

Since we are only interested in determining whether the scale $M_P/\xi$ has a physical meaning, we shall momentarily leave out all terms which are suppressed by $M_P/\sqrt{\xi}$ or a larger scale, \ie we consider
\bea  
\label{actionHiggsLeadingShort}
&&S\simeq  -\f 1 2\int\diff^4x \Big[(\p_\mu \chi)^2+(\p_\mu\sigma_a)^2+ \f{\xi}{M_P}\sqrt{6}\p_\mu\chi\p^\mu\sigma_a^2+\f{\xi^2}{M_P^2}\f 3 2  (\p_\mu\sigma_a^2)^2 \Big]\ .
\eea
We can rewrite this action in a compact sigma-model form, in complete analogy with~\eqref{eq:sigma_model_form}. Obviously now the internal field-space metric is a $4\times 4$ (nondegenerate) matrix that reads
\be
G_{IJ}(\sigma_1,\sigma_2,\sigma_3)=\begin{pmatrix}
	1 & \sqrt{6}\f{\xi}{M_P}\sigma_1& \sqrt{6}\f{\xi}{M_P}\sigma_2& \sqrt{6}\f{\xi}{M_P}\sigma_3 \\
	\sqrt{6}\f{\xi}{M_P}\sigma_1& 1+6\f{\xi^2}{M_P^2}\sigma_1^2 & 6\f{\xi^2}{M_P^2}\sigma_1 \sigma_2 &6\f{\xi^2}{M_P^2}\sigma_1 \sigma_3\\
	\sqrt{6}\f{\xi}{M_P}\sigma_2&6\f{\xi^2}{M_P^2}\sigma_1 \sigma_2& 1+6\f{\xi^2}{M_P^2}\sigma_2^2 & 6\f{\xi^2}{M_P^2}\sigma_2 \sigma_3\\
	\sqrt{6}\f{\xi}{M_P}\sigma_3&6\f{\xi^2}{M_P^2}\sigma_1 \sigma_3& 6\f{\xi^2}{M_P^2}\sigma_2\sigma_3& 1+6\f{\xi^2}{M_P^2} \sigma_3^2 
\end{pmatrix} \ ,
\ee
with $I,J=1,\ldots, 4$.
Using 
\begin{equation*}
	\kappa^I_{~JKL}=\p_K \gamma^I_{LJ}-\p_L \gamma^I_{KJ}+\gamma^I_{KM}\gamma^M_{LJ}-\gamma^I_{LM}\gamma^M_{KJ} \ ,~~~\gamma^I_{MN} = \f 1 2 G^{IJ}\l( \p_M G_{J N} +\p_N G _{MJ}-\p_{J} G_{MN}\r) \ ,
\end{equation*}
we find that $\kappa^I_{~JKL}=0$, meaning that the manifold is flat. Thus, the model~\eqref{actionHiggsLeadingShort} represents a free theory and the scale $M_P/\xi$ does not correspond to a cutoff. This agrees with the conclusions of \cite{Mikura:2021clt}.

The advantage of studying field space curvature is that it corresponds to a straightforward and widely applicable analysis. However, the calculations are involved and do not reveal in which channel unitarity is violated. Therefore, our investigation which relies on field redefinitions, such as shown in \Eq~\eqref{fieldRedefinition}, can be regarded as complementary to that of \cite{Mikura:2021clt},  making the results more transparent and allowing to single out the field variables leading to a simple extraction of the cutoff value.

 \subsection{Analysis in angular coordinates}
 \label{ssec:angularAnalysis}
 We shall now discuss in details yet another way to arrive at the conclusion that $M_P/\xi$ does not represent an inflationary cutoff scale. To this end, instead of~\eqref{doubletCartesian}, we now consider the nonlinear parametrization for the Higgs
 \begin{equation} 
  \label{doubletAngular}
 	H = \frac{1}{\sqrt{2}} (\bar{h} + h) \ex^{i\f{ \pi_a \tau_a}{\bar h}}
 	\begin{pmatrix}
 		0\\
 		1
 	\end{pmatrix} \ ,
 \end{equation}
 where $\tau^a$, $a = 1,2,3$, are the Pauli matrices, and as before the $\pi_a$'s represent three real fields. We remark that, obviously, \Eq \eqref{doubletAngular} is not the only parametrization that fulfills our purposes.\footnote{For example, one could also use
  \begin{equation} \label{doubletAlternative}
 	H = \frac{1}{\sqrt{2}} (\bar{h} + h) \ex^{i \alpha}
 	\begin{pmatrix}
 		\sqrt{2}\,	\pi^+ \\
 		\sqrt{1-2 \pi^+ \pi^-}
 	\end{pmatrix} \;,
\end{equation}
where $\alpha$ is a real scalar and $\pi^\pm=\f{1}{\sqrt 2}(\pi_1\mp i\pi_2)$. Although nonstardard, an advantage of the parametrization~\eqref{doubletAlternative} would be that there are no interactions of $\alpha$ and $\pi^\pm$.}

Plugging~\eqref{doubletAngular} into the action~\eqref{eq:action_starting} and keeping terms which are at most quartic in the fields, we obtain 
\be
S=\int \diff^4x\left(\mathcal L_2 + \tilde{\mathcal L}_3 +\tilde{\mathcal L}_4\right)\ ,
\ee
where the quadratic part is given as before by \Eq~\eqref{eq:L2}. In contrast, the tildes indicate that the parts of the Lagrangian which are cubic and quartic in the perturbations are different from---and actually quite simpler than---their Cartesian counterparts appearing in \Eqs~\eqref{eq:L3} and \eqref{eq:L4}, and read
\bea
&\tilde{\mathcal L}_3= -\f{1}{\bar h\bar \Omega^4}\left[\left(-\f{\xi\bar h^2}{M_P^2}+\f{6\xi^2\bar h^2}{M_P^2\bar\Omega^2}\left(1-\f{\xi\bar h^2}{M_P^2}\right)\right)h (\p_\mu h)^2+h (\p_\mu \pi_a)^2 \right] \ ,~&\\
&\tilde{\mathcal L}_4=-\f{1}{2\bar h^2\bar \Omega^6}\Bigg[ \f{\xi\bar h^2}{M_P^2}\left(3(1+6\xi)\bar \Omega^2-4(1+21\xi)+\f{72\xi}{\bar \Omega^2}\right)h^2(\p_\mu h)^2 \quad&\nonumber\\
&\qquad\qquad\qquad+ \left(1-\f{3\xi\bar h^2}{M_P^2}\right)h^2(\p_\mu\pi_a)^2 -\f{\bar\Omega^4}{3}\left(\pi_a^2(\p_\mu \pi_b)^2-\f 1 4 (\p_\mu\pi_a^2)^2\right)\Bigg] \ .&
\eea
 We make the kinetic terms canonical by using the same rescaling as in the previous section, see~(\ref{eq:field_resc}). This leads to the Lagrangian $\mathcal L_2$ shown in \Eq~\eqref{eq:can_kin_L2}, as well as
  \bea
&\tilde{\mathcal L}_3=\f{1}{\bar h\sqrt{\bar\Omega^2+\f{6\xi^2\bar h^2}{M_P^2}}}\left[\f{\xi\bar h^2}{M_P^2}\left(1-\f{6\xi}{\bar\Omega^2+\f{6\xi^2\bar h^2}{M_P^2}}\right)\chi(\p_\mu \chi)^2-\chi (\p_\mu \sigma_a)^2 \right] \ ,~&\\
&\tilde{\mathcal L}_4=-\f{1}{2\bar h^2}\Bigg[ \f{\xi\bar h^2}{M_P^2}\f{3(1+6\xi)\bar \Omega^4-4(1+21\xi)\bar \Omega^2+72\xi}{\left(\bar \Omega^2+\f{6\xi^2\bar h^2}{M_P^2}\right)^2}\chi^2(\p_\mu \chi)^2 \qquad\qquad\qquad\quad&\nonumber\\
&\qquad\quad\,+ \f{1-\f{3\xi\bar h^2}{M_P^2}}{\bar \Omega^2+\f{6\xi^2\bar h^2}{M_P^2}}\chi^2(\p_\mu\sigma_a)^2 -\f{\bar\Omega^2}{3}\left(\sigma_a^2(\p_\mu \sigma_b)^2-\f 1 4 (\p_\mu\sigma_a^2)^2\right)\Bigg] \ .&
\eea
  
   Considering the limit $\bar h\gg M_P/\sqrt{\xi}$,  it is easy to see that in terms of the canonically normalized fields, the action during inflation becomes
\bea
S=-\f 1 2\int \diff^4 x \Bigg[(\p_\mu\chi)^2+ (\p_\mu\sigma_a)^2-\f{\xi}{3M_P^2}\left(\sigma_a^2(\p_\mu \sigma_b)^2-\f 1 4 (\p_\mu\sigma_a^2)^2\right)+\ldots \Bigg]\ ,~~
\eea
where the ellipses stand for the operators suppressed by the Planck mass as well $\xi\bar h^2$; the effect of the latter during inflation is completely negligible.  In the variables~(\ref{doubletAngular}), the smallest energy scale that exists is $M_P/\sqrt{\xi}$, which only appears due to the self-interactions of the $\sigma^a$'s. We conclude that the lower bound on the cutoff is the same as the one derived in Cartesian coordinates, see \Eq~(\ref{eq:true_cutoff}).
Moreover, the interactions of $h$ with any of the three other modes is heavily suppressed by the scales
\begin{equation}
\sqrt{\xi} \bar{h} \ , \qquad \frac{\xi\bar{h}^2}{M_P} \  .
\end{equation}
Thus, $h$ decouples completely from the $\sigma_a$ modes in the limit $\bar h \rightarrow \infty$.

\section{Palatini Higgs inflation}
\label{sec:PalatiniHiggs}

Before moving to the conclusions, let us briefly discuss what happens if one replaces the metric formulation of gravity by the Palatini one. Then, instead of~(\ref{eq:action_starting}), the action simplifies considerably and contains only one term: 
\be
\label{eq:action_starting_2}
S= -\int \diff^4x\, \f{1}{\Omega^2}\p_\mu H^\dagger\p^\mu H   \ .
\ee
Notice the absence of the dimension-six operator, the origin of which is the inhomogeneous transformation of the Ricci tensor $R_{\mu\nu}$---in the Palatini formulation, $R_{\mu\nu}$ is inert.
Due to this fact, the fictitious scale $M_P/\x$ does not appear in the action for the excitations, irrespectively of whether the Cartesian or angular parametrization for the would-be NG modes is employed. 

\subsection{Analysis in Cartesian coordinates}

Just like in the metrical formulation of HI, we start by working with the linear, Cartesian, parametrization for the would-be NG bosons, see~(\ref{doubletCartesian}). Plugging that into~(\ref{eq:action_starting_2}), expanding in $h$ and $\pi_a$,  and keeping terms at most quartic in these fields, we find
\be
S = \int \diff^4x\l(\mathscr L_2+\mathscr L_3+\mathscr L_4\r)  \ ,
\ee
with 
\bea
\label{eq:kin_terms_palat_cartes}
&&\mathscr L_2 = - \f{1}{2\bar \Omega^2}\left((\p_\m h)^2 +(\p_\m\pi_a)^2\right)   \ ,\\
&&\mathscr L_3= \f{\x\bar h}{M_P^2\bar \Omega^4}h\left[ (\p_\mu h)^2+ (\p_\mu \pi_a)^2 \right] \ ,\\
&&\mathscr L_4 = \f{\x}{2M_P^2\bar \Omega^6}\Bigg[\left(h^2\left(1-\f{3\xi\bar h^2}{M_P^2}\right)+\bar \Omega^2\pi_a^2\right)\left((\p_\mu h)^2+(\p_\mu\pi_b)^2\right)\Bigg] \ . 
\eea
We notice from~(\ref{eq:kin_terms_palat_cartes}) that all kinetic terms become canonical once the fields are rescaled with $\bar\Omega$, \ie 
\be \label{rescalingPalatini}
h = \bar\Omega \chi \ ,~~~\pi_a = \bar\Omega \sigma_a \ . 
\ee
This leads to 
\bea
\label{eq:kin_terms_palat_canonical_cartes}
&&\mathscr L_2=  - \f{1}{2}\left((\p_\m \chi)^2 +(\p_\m\sigma_a)^2\right)   \ ,\\
\label{eq:palat_canonical_cartes_3}
&&\mathscr L_3= \f{\x\bar h}{M_P^2\bar \Omega}\chi\left[ (\p_\mu \chi)^2+ (\p_\mu \sigma_a)^2 \right] \ ,\\
\label{eq:palat_canonical_cartes_4}
&&\mathscr L_4 = \f{\x}{2 M_P^2\bar \Omega^2}\Bigg[\left(\chi^2\left(1-\f{3\xi\bar h^2}{M_P^2}\right)+\bar \Omega^2\sigma_a^2\right)\left((\p_\mu \chi)^2+(\p_\mu\sigma_b)^2\right)\Bigg]  \ . 
\eea
The above makes clear that the lowest scale appearing is proportional to $M_P/\sqrt{\x}$, irrespectively of the inflationary limit. Nevertheless, we are interested in the energy domain $\bar h\gg M_P/\sqrt{\xi}$, where the action in terms of the canonically normalized variables becomes 
\bea
&&
S=-\f 1 2\int \diff^4 x \Bigg[(\p_\mu\chi)^2+ (\p_\mu\sigma_a)^2-\f{2\sqrt{\xi}}{M_P}\chi \left((\p_\m \chi)^2+(\p_\mu\sigma_a)^2\right)\nonumber\\
&&
\qquad\qquad\qquad\qquad\quad+ \f{ \x}{M_P^2}\left(3\chi^2-\sigma_a^2\right)\left((\p_\m \chi)^2+(\p_\mu\sigma_b)^2\right)+\ldots \Bigg] \ .
\eea
Here we only kept the leading higher-dimensional operators suppressed by powers of $M_P/\sqrt{\xi}$. 

\subsection{Analysis in angular coordinates}

Let us now work with the non-linear parametrization of the Higgs doublet in terms of $SU(2)$ generators given in \Eq \eqref{doubletAngular}. Then from~(\ref{eq:action_starting_2}), we obtain the action up to terms quartic in the perturbations $h$ and $\pi_a$:
\be
\label{eq:action_perts_palat}
    S= \int \diff^4x \Big(\mathscr L_2+\widetilde{\mathscr L}_3+\widetilde{\mathscr L}_4 \Big) \ ,
\ee
with $\mathscr L_2$ given by~(\ref{eq:kin_terms_palat_cartes}), while 
\bea
\label{eq:kin_terms_palat}
&&\widetilde{\mathscr L}_3= -\f{h}{\bar h\bar \Omega^4}\left[-\f{\xi\bar h^2}{M_P^2}(\p_\mu h)^2+ (\p_\mu \pi_a)^2 \right] \ ,\\
&&\widetilde{\mathscr L}_4 = -\f{1}{2\bar h^2\bar \Omega^6}\Bigg[h^2\left(1-\f{3\xi\bar h^2}{M_P^2}\right)\left(-\f{\xi \bar h^2}{M_P^2}(\p_\mu h)^2+(\p_\mu\pi_a)^2\right)\nonumber\\
&&\qquad\qquad\qquad\qquad\qquad\qquad\qquad-\f{\bar\Omega^4}{3}\left(\pi_a^2(\p_\mu \pi_b)^2-\f 1 4 (\p_\mu\pi_a^2)^2\right)\Bigg] \ . 
\eea
Notice that contrary to what happens in metric HI, the quartic piece of the resulting action in terms of the angular variables is actually more involved than in the Cartesian ones.

Again, we make the kinetic terms canonical by using~(\ref{rescalingPalatini}), and end up with
\bea
\label{eq:kin_terms_palat_canonical}
&&\widetilde{\mathscr L}_3= -\f{\chi}{\bar h\bar \Omega}\left[-\f{\xi\bar h^2}{M_P^2} (\p_\mu \chi)^2+ (\p_\mu \sigma_a)^2 \right] \ ,\\
&&\widetilde{\mathscr L}_4 = -\f{1}{2\bar h^2\bar \Omega^2}\Bigg[\chi^2\left(1-\f{3\xi\bar h^2}{M_P^2}\right)\left(-\f{\xi \bar h^2}{M_P^2}(\p_\mu \chi)^2+(\p_\mu\sigma_a)^2\right)\nonumber\\
&&\qquad\qquad\qquad\qquad\qquad\qquad\qquad
-\f{\bar\Omega^4}{3}\left(\sigma_a^2(\p_\mu \sigma_b)^2-\f 1 4 (\p_\mu\sigma_a^2)^2\right)\Bigg] \ . 
\eea
Like in~(\ref{eq:palat_canonical_cartes_3}) and~(\ref{eq:palat_canonical_cartes_4}), the leading operators are suppressed by the scale $M_P/\sqrt{\x}$. This becomes apparent from the action in the inflationary limit,
\bea
&&
S\simeq-\f 1 2\int \diff^4 x \Bigg[(\p_\mu\chi)^2+ (\p_\mu\sigma_a)^2 -\f{2\sqrt{\xi}}{M_P}\chi (\p_\m \chi)^2\nonumber\\
&&
\qquad+\f{\x}{M_P^2}\left( 3\chi^2 (\p_\m\chi)^2 -\f 1 3 \left(\sigma_a^2(\p_\mu \sigma_b)^2-\f 1 4 (\p_\mu\sigma_a^2)^2\right) \right) + \ldots
\Bigg] \ ,
\eea
where as before we only kept the leading higher-dimensional operators. Just like in the metric case, there are also terms suppressed by $M_P$ as well as $M_P/\xi \bar h^2$, and the exponential parametrization given in \Eq \eqref{doubletAngular} is advantageous because $h$ decouples from the angular modes $\sigma_a$ in the limit of a large background field. As already discussed after \Eq~\eqref{eq:cutoffInflationUnitary}, we conclude that the inflationary cutoff in Palatini Higgs inflation coincides with the metric scenario and scales as $M_P/\sqrt{\xi}$.

\section{Conclusion}
\label{sec:conclusion}

When studying an inflationary model, it is important to make sure that the cutoff scale, above which perturbation theory breaks down, exceeds all relevant energies. In the present paper, we investigated this question for the proposal that the Higgs boson acted as the inflaton. As in the original model, we used the metric formulation of General Relativity. The phenomenological viability of this scenario requires the presence of a large non-minimal coupling between the Higgs doublet and gravity. It has been long known that the perturbative unitarity of the theory is violated at energies well below the Planck scale. Nevertheless, the fact that the cutoff is significantly higher during inflation than in vacuum ensures the validity of effective field theory and the consistency of predictions derived from it. The goal of the present short note was to confirm this result using a new, and arguably simpler, method.

The inflationary value of the cutoff is usually  derived in the unitary gauge. There, the scalar \& gauge spectra of the theory comprise only the physical Higgs and vector bosons. The failure of perturbation theory in this setting is well understood and has been studied in detail: it is due to the inability of the Higgs to unitarize scattering processes involving longitudinal gauge bosons. In turn, this is a result of the non-minimal coupling $\xi$ of the Higgs field to gravity, which practically disentangles the self-interactions of the gauge bosons from their couplings with the Higgs background and excitations. An inspection of the relevant pieces of the action reveals that the effective mass of the gauge bosons during inflation, and thus the cutoff, is proportional to $M_P/\sqrt{\xi}$.

Of course, it should be possible to obtain this result without employing unitary gauge. In this case, the problem is already apparent in the scalar sector of the theory and more  specifically in the (self-)interactions of would-be NG modes, which are now present in the action. This is expected, since after all, these are the longitudinal components of the non-abelian vectors in unitary gauge. However, if the Cartesian parametrization for the Higgs doublet is used,  the action for the excitations features certain higher-dimensional operators that are only suppressed by the significantly lower scale $M_P/\xi$. This could create the (false) impression that the inflationary cutoff might in fact lie below $M_P/\sqrt{\xi}$. 

In the present paper, we pointed out a way to show that this is not the case, without any need to perform lengthy and potentially error-prone computations of amplitudes. Instead, we demonstrated that there exists a simple field redefinition that removes all those problematic operators. This analysis made apparent that the true cutoff corresponds to $M_P/\sqrt{\xi}$ whereas the presence of the low scale $M_P/\xi$ is an artifact of an unsuitable choice of field variables. Moreover, we showed that one can arrive at this conclusion in a more direct way. If one uses the usual exponential parametrization for the NG modes, such seemingly dangerous terms are absent from the beginning and it is not necessary to perform any field redefinitions (apart from an obvious rescaling in order to make the kinetic terms canonical). This alternative representation of the NG modes can potentially be useful for the study of Higgs inflation both in the metric and Palatini scenarios.

\section*{Acknowledgments} 

This work was supported by the ERC-AdG-2015 grant 694896. We thank Andrey Shkerin and Inar Timiryasov for discussions and comments on the manuscript. 

\bibliographystyle{utphys}
\bibliography{Refs}{}

\end{document}